\documentclass[fleqn,usenatbib]{mnras}

\usepackage{newtxtext,newtxmath}

\usepackage[T1]{fontenc}

\DeclareRobustCommand{\VAN}[3]{#2}
\let\VANthebibliography\thebibliography
\def\thebibliography{\DeclareRobustCommand{\VAN}[3]{##3}\VANthebibliography}

\usepackage{graphicx}	
\usepackage{amsmath}

\usepackage[acronym]{glossaries}
\makeglossaries
\newacronym{GW}{GW}{gravitational wave}
\newacronym{NS}{NS}{neutron star}
\newacronym{BH}{BH}{black hole}
\newacronym{BBH}{BBH}{binary black hole}
\newacronym{BNS}{BNS}{binary neutron star}
\newacronym{BHNS}{BHNS}{black hole-neutron star}
\newacronym{EoS}{EoS}{equation of state}
\newacronym{EM}{EM}{electromagnetic}
\newacronym{sGRB}{sGRB}{short Gamma-Ray Burst}
\newacronym{ISCO}{ISCO}{Innermost Stable Circular Orbit}
\newacronym{CTT}{CTT}{Conformal Transverse-Traceless}
\newacronym{CTS}{CTS}{Conformal Thin Sandwich}
\newacronym{xCTS}{xCTS}{extended Conformal Thin Sandwich}
\newacronym{ID}{ID}{initial data}
\newacronym{PDE}{PDE}{partial differential equation}
\newacronym{NR}{NR}{numerical relativity}
\newacronym{GR(M)HD}{GR(M)HD}{General Relativistic (Magneto) Hydrodynamics}
\newacronym{HRSC}{HRSC}{High-Resolution Shock-Capturing}

\usepackage{siunitx}
\DeclareSIUnit{\erg}{erg}
\DeclareSIUnit{\Mass}{\mathit{M}}
\DeclareSIUnit{\c}{\mathit{c}}
\DeclareSIUnit{\dyn}{dyn}
\DeclareSIUnit{\Gauss}{G}
\DeclareSIQualifier{\Sun}{\ensuremath{\odot}}
\DeclareSIQualifier{\disk}{disk}
\DeclareSIQualifier{\BH}{BH}
\DeclareSIQualifier{\NS}{NS}
\DeclareSIUnit\clight{\text{\ensuremath{c}}}

\usepackage[capitalise,noabbrev]{cleveref}

\title{Signatures of Low Mass Black Hole-Neutron Star Mergers}

\author[R. Matur, I. Hawke and N. Andersson]{
Rahime Matur,$^{1}$\thanks{E-mail: r.matur@soton.ac.uk}
Ian Hawke,$^{1}$
Nils Andersson$^{1}$
\\
$^{1}$Mathematical Sciences and STAG Research Centre University of Southampton, Southampton SO17 1BJ, UK\\
}

\date{Accepted XXX. Received YYY; in original form ZZZ}

\pubyear{\the\year{}}

\begin{document}
\glsdisablehyper
\label{firstpage}
\pagerange{\pageref{firstpage}--\pageref{lastpage}}
\maketitle

\begin{abstract}
The recent observation of the GW230529 event indicates that black hole-neutron star binaries can contain low-mass black holes. Since lower mass systems are more favourable for tidal disruption, such events are promising candidates for multi-messenger observations. In this study, we employ five finite-temperature, composition-dependent matter equations of state and present results from ten 3D general relativistic hydrodynamic simulations for the mass ratios $q = 2.6$ and $5$. Two of these simulations target the chirp mass and effective spin parameter of the GW230529 event, while the remaining eight contain slightly higher-mass black holes, including both spinning ($a_{BH} = 0.7$) and non-spinning ($a_{BH} = 0$) models. We discuss the impact of the equation of state, spin, and mass ratio on black hole-neutron star mergers by examining both gravitational-wave and ejected matter properties. For the low-mass ratio model we do not see fast-moving ejecta for the softest equation of state model, but the stiffer model produces on the order of $10^{-6}M_\odot$ of fast-moving ejecta, expected to contribute to an electromagnetic counterpart. Notably, the high-mass ratio model produces nearly the same amount of total dynamical ejecta, but yields $52$ times more fast-moving ejecta than the low-mass ratio system. In addition, we observe that the black-hole spin tends to decrease the amount of fast-moving ejecta while increasing significantly the total ejected mass.
Finally, we note that the disc mass tends to increase as the neutron star compactness decreases.
\end{abstract}
\begin{keywords}
stars: neutron -- stars: black holes --  gravitational waves --  hydrodynamics
\end{keywords}

\section{Introduction}

The first gravitational-wave event, GW150914 \citep{firstbbh}  detected in 2015, corresponded to the late inspiral and merger of a binary black-hole system. Two years later, ground-based detectors observed the first multi-messenger event from a binary neutron star merger, GW170817 \citep{2017PhRvL.119p1101A, 2017ApJ...848L..13A}, marking another important breakthrough in the field. In particular, by combining gravitational wave and electromagnetic observations, the tidal deformability parameter could be constrained, providing important insight into the equation of state of matter at supranuclear densities. 

Following these achievements, the first two black hole-neutron star mergers, GW200105 and GW200115 \citep{2021ApJ...915L...5A}, were detected by LIGO and Virgo in 2020, but no associated electromagnetic counterparts were observed. Very recently, the most symmetric black hole-neutron star merger event (in terms of mass), GW230529 \citep{2024arXiv240404248T}, was observed by LIGO. This system contains the smallest black hole seen so far, with a mass of $3.6^{+0.8}_{-1.2} \unit{\Mass\Sun}$,  placing it in the so-called mass gap. Due to the low black-hole mass, the GW230529 event would be expected to exhibit tidal disruption, in turn producing an electromagnetic counterpart. However, since this event was only detected by a single detector (LIGO Livingston), the sky position was not well constrained, making it impossible to find any accompanying electromagnetic signal. Nevertheless, GW230529 is highly significant as it demonstrates that black hole-neutron star mergers can be strong candidates for multi-messenger observations. In fact, events involving low-mass black holes may be surprisingly common. 

Theoretical work suggests that the tidal disruption of the neutron star depends on three important parameters: the mass ratio ($q = M_{BH}/M_{NS}$), the neutron star compactness ($\mathcal{C} = GM_{NS}/(c^{2}R_{NS})$), and the dimensionless spin parameter of the black hole ($a_{BH} = cJ_{BH}/(GM_{BH}^{2})$, restricted to be in the range $-1<a_{BH}<1$) (see~\cite{foucartreview,kyutokureview} for detailed discussion). If the mass ratio is small and/or the dimensionless spin parameter of the black hole is high and/or the compactness of the neutron star is small, the neutron star is expected to be tidally disrupted. 

An understanding of black hole-neutron star mergers requires numerical relativity simulations. These simulations allow us to extract gravitational-wave signals, essential for distinguishing signals from binary neutron star and binary black hole mergers. Additionally, we can infer the amount of ejected matter and its composition. The outcome of such simulations can then---at a post-processing stage---be used to investigate short gamma-ray bursts, r-process nucleosynthesis, and produce kilonova light curves. This, in turn, allows us to constrain the equation of state of the neutron star, determine the dimensionless spin parameter of the black hole, and understand the contribution of black hole-neutron star mergers to the abundance of heavy elements in the universe. 

The first fully general relativistic black hole-neutron star merger simulations used $\Gamma$-law equations of state~\citep{frankheadon, shibatafirst, duezfirst, firstetienne}. Simulations involving realistic finite-temperature, composition-dependent equations of state followed more recently, see~\cite{firsttabulated, firsttab2,2017CQGra..34d4002F}. Similarly, the effect of the equation of state has been investigated using various piecewise polytropic and $\Gamma$-law models in~\cite{gammalaw1, gammalaw6, gammalaw7, gammalaw3, gammalaw2, gammalaw4, gammalaw5, tsaobhns}, and, only relatively recently, using different finite-temperature, composition-dependent equations of state in~\cite{kyutoku2018,foucarteos, 2021MNRAS.506.3511M, 2023PhRvD.107l3001H}.

In this study, we perform ten numerical relativity simulations using five finite-temperature, composition-dependent tabulated equations of state. We run eight simulations for a high mass ratio ($q = 5$) and two for a low mass ratio ($q \simeq 2.6$). The chirp mass and the effective spin of the latter simulations are consistent with the GW230529 event. The aim is to estimate how the equation of state, mass ratio, and black-hole spin impact on the merger dynamics, the mass and composition of the total and fast-moving ejecta, the properties of the disc and remnant, etc. Furthermore, we aim to identify which parameters most significantly influence these dynamics and what the results would be if we analysed the behaviour in detail for an observed system like GW230529.

The paper is organized as follows: In \cref{sec:numerics}, we explain the numerical setup. In \cref{sec:results}, in addition to analysing gravitational-wave signals and ejected matter properties, we present the possibility of distinguishing \gls{EoS} effects. Finally, in \cref{sec:conclusion}, we summarize our findings. We used geometrized units, setting $G$ and $c$ equal to $1$ unless noted otherwise.

\begin{figure}
	\includegraphics[width=\columnwidth]{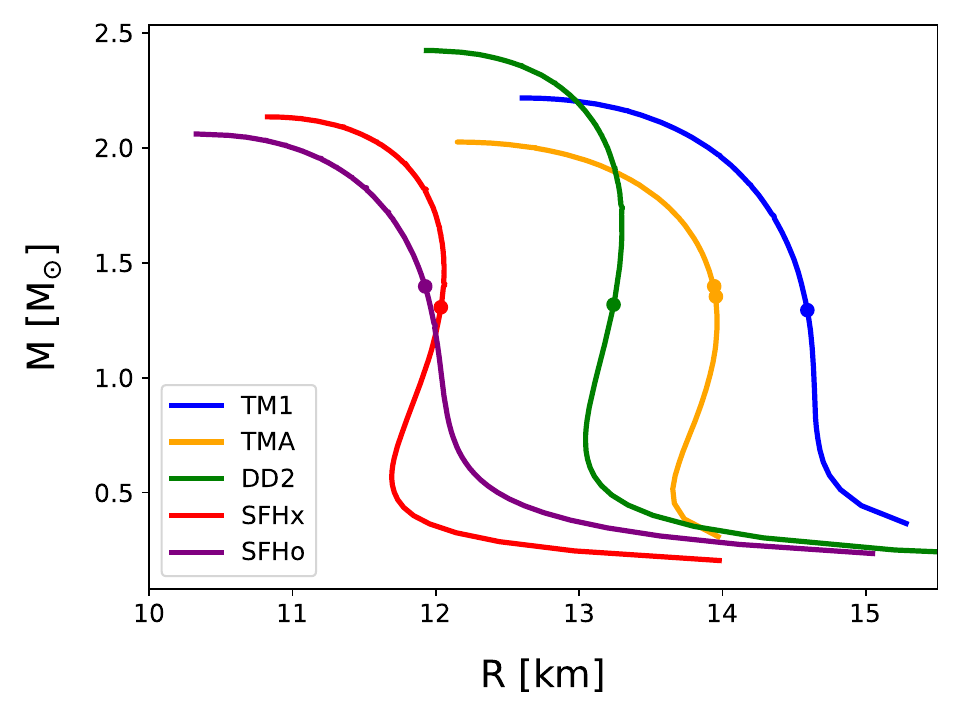}
    \caption{Mass-radius relation for our chosen set of equations of state. Circles indicate the masses of the models used in this study. Two configurations have been modelled with the DD2 \gls{EoS}, therefore there are two filled circles on that curve.}
    \label{fig:mass_radius}
\end{figure}

\section{Numerical Setup}\label{sec:numerics}

We have performed ten \gls{NR} simulations for \gls{BHNS} mergers. All \gls{ID} was generated using the \texttt{Fuka} ID solver (\cite{2021PhRvD.104b4057P}), which uses the extended-conformal thin sandwich method (\cite{1999PhRvL..82.1350Y, 2003PhRvD..67d4022P}).
We used \texttt{WhiskyTHC}~\citep{2012A&A...547A..26R, 2014CQGra..31g5012R, 2014MNRAS.437L..46R, 2015ASPC..498..121R}, built on top of Cactus~\citep{cactus}, for the hydrodynamical evolution. We modelled the \gls{NS} as a perfect fluid through the conservation equations,
\begin{equation}
    \nabla_{\mu} T^{\mu\nu} = 0, \\
    \nabla_{\mu} (\rho u^{\mu}) = 0
\end{equation}
where the stress-energy tensor $T_{\mu\nu}$ is given by, 
\begin{equation}
    T_{\mu\nu} = \rho h u_\mu u_\nu + p g_{\mu\nu}
\end{equation}
with $\rho$, $u^{\mu}$, $h$, $p$ and $g_{\mu\nu}$ representing the rest-mass density, four-velocity, enthalpy, pressure of the fluid, and the metric tensor, respectively. \texttt{WhiskyTHC} writes these conservation equations in a flux-conservative form known as the Valencia formulation (\cite{1991PhRvD..43.3794M, 1997ApJ...476..221B, 1999astro.ph.11034I}) to accurately evolve the fluid variables in the presence of shocks (\cite{richest}).

\texttt{WhiskyTHC} further uses a \gls{HRSC} scheme to divide the domain into cells and calculate the primitive variables on cell interfaces. We also use the Local Lax Friedrichs flux splitting method from \cite{llf}. We use the fourth-order Runge-Kutta (RK4) method for time integration and set the Courant-Friedrichs-Lewy (CFL) condition to 0.15. The atmosphere temperature and rest-mass density are set to 
$\qty{0.02}{\mega\eV}$ and $\qty{6.2e3}{\gram\per\centi\metre\cubed}$ ($\rho_{atmo}$), respectively.

\begin{table}\label{tab:eos}
	\centering
	\caption{The properties of the stellar models used in this study. Columns provide the maximum supported mass for a single TOV model, the radius of the \gls{NS} for $\qty{1.4}{\Mass\Sun}$, and the compactness of the \gls{NS} for the same mass. \glspl{EoS} are listed from the stiffest to the softest.}
	\label{tab:eostable}
	\begin{tabular}{@{}lSSS@{}} 
  		\hline
		Name & $\unit{\Mass}_{max} [\unit{\Mass\Sun}]$ & $R_{1.4} [\unit{\kilo\metre}]$ & $\mathcal{C}_{1.4}$\\
		\hline
        TM1&2.21&14.5&0.143\\
        TMA&2.02&13.8&0.150\\
        DD2&2.42&13.2&0.157\\
        SFHx&2.13&12.0&0.172\\
        SFHo&2.06&11.9&0.174\\
	    \hline
	\end{tabular}
\end{table}

We use \texttt{CTGamma} (\cite{ctgamma}), which employs the \texttt{z4c} formulation (\cite{z4c}) of the Einstein Equations, for the spacetime evolution. The \texttt{z4c} formulation introduces constraint-damping parameters, which helps reduce the constraint violation and therefore provides more stable evolutions. 

The choice of gauge conditions has crucial impact on the stability
of the numerical evolution of the spacetime. We use the 1+log slicing and the Gamma Driver shift condition to evolve the lapse function and the shift vector. We also set the constraint damping coefficients for \texttt{z4c}; $\kappa_{1}$ and $\kappa_{2}$ to $0$ and $0.02$, respectively. The damping coefficient for the shift evolution is set to $2/\unit{\Mass}_{tot}$, where $\unit{\Mass}_{tot}$ is the total gravitational mass of the system.

In all simulations, we employ reflection symmetry on the lower $z$ axis, therefore the domain extends to $(x,y,z) = (2835,2835,1418) [\unit{\kilo\metre}]$. We use \texttt{Carpet} adaptive mesh refinement (AMR) driver (\cite{carpet}) of \texttt{Cactus}, set 8 refinement levels for all simulations and re-grid every 128 iterations to update the refinement levels. We use \texttt{AHFinderDirect} \citep{ahfinder} to locate the apparent horizon, \texttt{PunctureTracker} to track the \gls{BH} and \texttt{CarpetTracker} to update the refinement levels to move them as \gls{BH} moves. The \gls{NS} is being tracked with \texttt{VolumeIntegralGRMHD}. These thorns are part of \href{https://einsteintoolkit.org/}{\texttt{Einstein Toolkit}} \citep{etcode} for which we use the \textit{Sophie Kowalevski} release \citep{2022zndo...7245853H}.  

We analyse our results using the \texttt{PostCactus} library, see~\cite{kastaun_pycactus_2021}, \cite{petergw, petergw2} and \href{https://bitbucket.org/dradice/scidata/}{Scidata}. The \gls{GW} strain is computed from the double time integration of the Weyl scalar ($\Psi_{4}$) using fixed-frequency integration \citep{ffi}.  

For the ejected matter, we determine unbound matter using the geodesic and Bernoulli criteria \citep{kastaunejectedmass}, as explained  in~\cite{foucartejectedmass}. According to the geodesic criterion, a fluid element is considered unbound if its time component of the four-velocity is smaller than $-1$,
\begin{equation}
    u_{t} < -1
\end{equation}
assuming it follows a geodesic. The  Bernoulli criterion instead  assumes the fluid element is unbound if the multiplication of the enthalpy and the time component of the four-velocity is less than $-1$,
\begin{equation}
    hu_{t} < -1
\end{equation}
We consider two mass ratios ($q = 2.06$ and $q = 5$) and employ five finite-temperature, composition-dependent equations of state. The properties and mass-radius relations for these \glspl{EoS} are presented in \cref{tab:eostable} and in \cref{fig:mass_radius}, respectively. All \glspl{EoS} are available from \href{https://stellarcollapse.org}{StellarCollapse} \citep{stellarcollapse}. 

In the first part of our investigation, we focus on the GW230529 event. We perform two simulations with a mass ratio of $q=2.6$, where the individual gravitational masses are taken to be $\unit{\Mass\NS} = \qty{1.4}{\Mass\Sun}$ and $\unit{\Mass\BH} = \qty{3.6}{\Mass\Sun}$. In this case, the chirp mass is $\qty{1.91}{\Mass\Sun}$, which falls within the parameter range for the GW230529 observation, where the chirp mass was $1.94_{-0.04}^{+0.04} M_\odot$. We model the two objects as non-spinning, with an effective spin corresponding to zero. The effective spin parameter of the GW230529 event was measured to be $-0.10_{-0.14}^{+0.12}$, which is consistent with our assumptions.

For the $q = 2.6$ cases, we choose two finite-temperature, composition-dependent \glspl{EoS}: \texttt{SFHo} \citep{2013ApJ...774...17S} and \texttt{TMA} \citep{2012ApJ...748...70H}. \texttt{SFHo} represents the softest \gls{EoS} (the smallest radius for the same gravitational mass) used in this study, while \texttt{TMA} represents a very stiff \gls{EoS} (larger radius for the same mass). Despite fixing the gravitational masses of the two \glspl{NS}, the baryon masses of the stars are different. Labelling each simulation by the mass ratio, the \gls{EoS} and the black hole spin,  our \texttt{Q2.6TMAa0} model ($q=2.6$ for the TMA \gls{EoS} with a non-spinning black hole), has a \gls{NS} with baryon mass of $1.6 M_{\odot}$, whereas it is $1.62 M_{\odot}$ in the \texttt{Q2.6SFHoa0} case. We calculated gravitational masses using \href{https://lorene.obspm.fr/}{Lorene}, \citep{gourgoulhonparis}). We set the numerical resolution to $\sim \qty{221}{\metre}$ for both models.

For the $q = 5$ mass ratio models, we fix the individual masses to $\unit{\Mass\NS}^{ADM} = 1.4$ and $\unit{\Mass\BH}^{grav} = \qty{7}{\Mass\Sun}$. For this mass ratio, we also fix the initial separation to $\sim \qty{60}{\kilo\metre}$. The only variables we changed are the initial dimensionless spin parameter of the \gls{BH} ($a_{BH} = 0$ and $a_{BH} = 0.7$) and the \gls{EoS} of the \gls{NS}. We perform eight \gls{NR} simulations for these cases, employing four finite-temperature, composition-dependent \glspl{EoS}: \texttt{TM1}, \texttt{TMA}, \texttt{DD2} (\cite{2012ApJ...748...70H}), and \texttt{SFHx} (\cite{2013ApJ...774...17S}). For these simulations, we fix the resolution to $\sim \qty{295}{\metre}$. 

While \texttt{WhiskyTHC} is a robust code extensively used in  \gls{BNS} merger studies, to the best of our knowledge this is the first study using it to simulate \gls{BHNS} mergers. Hence, we compare our \texttt{DD2a07} model with two previous studies. 
First, we compare to the results from~\cite{2017CQGra..34d4002F} which uses the same initial configuration as our model except for having neutrino cooling and non-zero inclination angle of the spin with respect to the orbital angular momentum. That study was performed using the \href{https://www.black-holes.org/code/SpEC.html}{SpEC} code, a robust software platform used in several studies of \gls{BHNS} mergers. In \gls{BHNS} mergers, if the \gls{BH} spin is not aligned with the orbital angular momentum, the possibility of tidal disruption of the \gls{NS} is decreased. This then leads to a decrease in the mass of the ejected matter \citep{2017CQGra..34d4002F, kyutokureview}. For the \texttt{DD2} \gls{EoS} \cite{2017CQGra..34d4002F} reported a total ejected mass of $\qty{0.014}{\Mass\Sun}$ while  our \texttt{DD2a07} model leads to $\qty{0.044}{\Mass\Sun}$. A difference of about a factor of 3 when the spin is aligned with respect to the orbital angular momentum does not seem unreasonable.
Of course, using two different codes with different numerical setups can also cause minor differences.

We also compare with the results of~\cite{foucarteos}. They investigate the same initial configuration as our \texttt{DD2a07} model, except for having a higher spin parameter ($a_{BH} = 0.9$) and also incorporating neutrino emission and absorption in the simulation. In this case we find good agreement for the mass of the remnant and the ejected matter. Their reported mass of ejected matter is  $\qty{0.06}{\Mass\Sun}$ which is slightly larger than the results for our \texttt{DD2a07} model. This minor difference is likely attributable to the higher spin. 

\section{Simulation Results}\label{sec:results}

We discuss the results from our simulations for the chosen two mass ratios separately. In the first case, for $q=2.6$ (inspired by GW230529), the main focus is on the impact of the equation of state. In the second case, for $q=5$, we also consider effects associated with the \gls{BH} spin.

\subsection{Impact of the EoS for \texorpdfstring{$q = 2.6$}{q=2.6} }

\begin{figure*}
\includegraphics[width=1.0\textwidth]{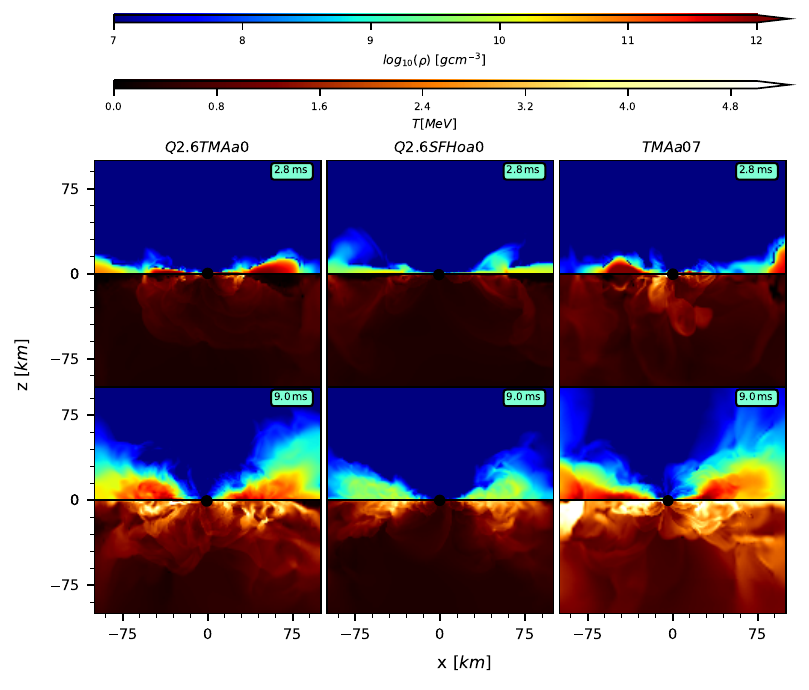}
\caption{Rest mass density and temperature plots in the $xz$ plane. The three columns represent three different models, from left to right: \texttt{Q2.6TMAa0}, \texttt{Q2.6SFHoa0}, and \texttt{TMAa07}, respectively. The top panels show the system $2.8$ ms after the merger, while the bottom panels show the same models at $9$ ms after the merger. In each individual panel, the top part represents the logarithm of the rest-mass density while the bottom part shows the temperature on a linear scale. It is evident that the angular distribution of all baryonic matter is in the equatorial plane at the first comparison time and extends to higher angles, although it remains very close to the equator.}
\label{figure*:gw2305292d}
\end{figure*}

\textit{Dynamics:} First of all, we compare the merger dynamics of the two low-mass \gls{BH} $q = 2.6$ models for which we present the rest-mass density and the temperature plots in the $xz$ plane at different times in \cref{figure*:gw2305292d}. We assume the \gls{NS} has been tidally disrupted if it leaves matter with a rest-mass density greater than $\qty{e6}{\gram\per\centi\metre\cubed}$ outside the \gls{BH}. Based on this criterion, even though we observe tidal disruption in both models, we see that the  \texttt{Q2.6TMAa0} model leaves more massive and dense matter around the \gls{BH}. During the evolution, while the temperature reaches up to $\qty{35}{\mega\eV}$, the disk remains relatively cold (see \cref{figure*:gw2305292d}).

\begin{figure}
\includegraphics[width=1.0\columnwidth]{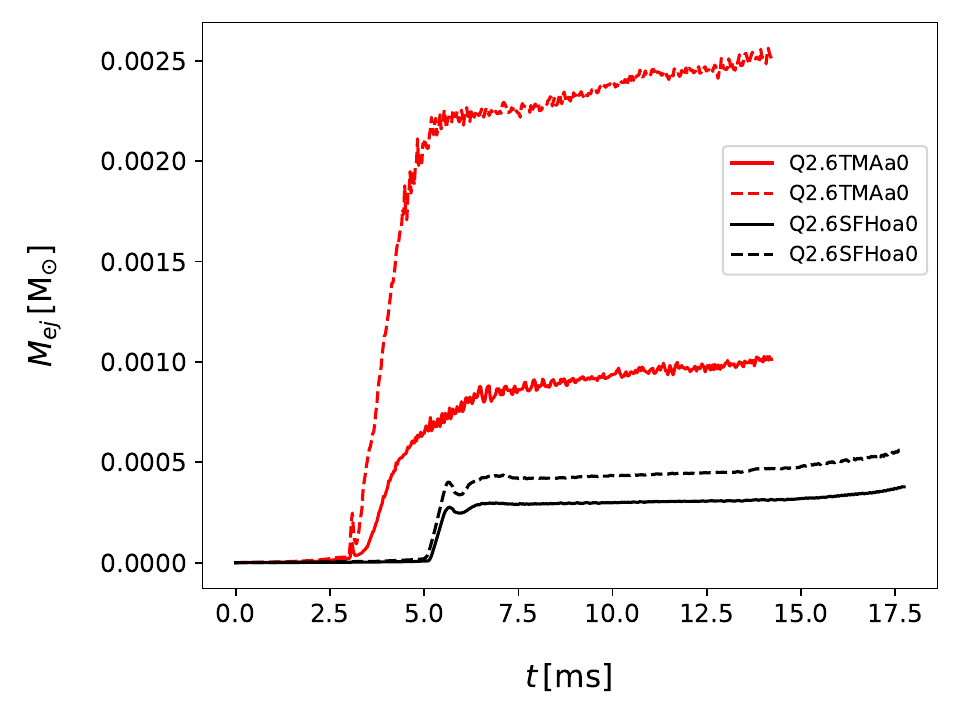}
    \caption{The evolution of the ejected matter calculated using the geodesic ($u_{t} < -1$) and Bernoulli ($hu_{t} < -1$) criteria. Solid lines show the ejected matter calculated by the geodesic criterion, while dashed lines show those calculated by the Bernoulli criterion. The results clearly show that stiffer \glspl{EoS} cause more ejecta compared to softer models.}
    \label{fig:massejection}
\end{figure}

\textit{Mass Ejection:} Next, we compare the properties of the ejected matter at two times: \qtylist{2.8;9}{\milli\second} after the merger. The ejected matter properties are extracted at a distance of about $ \qty{148}{\kilo\metre}$ and calculations are performed using the \texttt{Outflow} thorn. We present the total and fast-moving ejecta along with their composition and the amount of fallback material in~\cref{table:finalproperties} at \qty{2.8}{\milli\second} after the merger. \Cref{table:final2} shows the same quantities at \qty{9}{\milli\second} after the merger. We also present the evolution of the ejecta for these two models in~\cref{fig:massejection} for both the geodesic and the Bernoulli criterion.

\Cref{table:finalproperties} clearly illustrates the impact  of the \gls{EoS}. The total amount of ejected matter is more than twice as large as in the case of the stiffer \gls{EoS}. It is also more neutron-rich. This indicates that stiffer \glspl{EoS} are more favourable for r-process nucleosynthesis.

Previous hydrodynamical studies have shown that a portion of the total ejecta moves faster than \qty{0.6}{\clight}. This component, referred to as the fast-moving ejecta, is important for interpreting an \gls{EM} counterpart. This issue was discussed in~\cite{metzger, hotokezaka, radicemassejection}, and its application for distinguishing \gls{BNS} and \gls{BHNS} mergers was presented in~\cite{mostfastmoving}. While we do no observe a significant amount of fast-moving ejecta for the softer \gls{EoS} model, the fast-moving ejecta in the stiffer \gls{EoS} case is \qty{1.5e-6}{\Mass\Sun} at \qty{2.8}{\milli\second} after the merger. In general, the amount fast-moving ejecta for the stiffer \gls{EoS} is much higher than for softer ones, but, as explained in~\cref{section:highmass},  low mass ratio systems tend to produce less fast-moving ejecta.

At the second comparison time, (\qty{9}{\milli\second} after the merger), we note that the mass of the total ejecta remains nearly the same but the fast-moving component has increased. The composition of the total ejecta also remains the same, whilst the fast-moving ejecta becomes more neutron-rich (see \cref{table:final2}).

Finally, we compare our ejected matter properties with a similar recent study by \cite{foucartnew}. Their models have a mass ratio of $q = 3$ and they presented their results for different initial \gls{BH} spins. They used the \texttt{DD2} and \texttt{SFHo} \glspl{EoS}, which differ in compactness by $0.017$. Our results are consistent with their results.

\textit{Spiral Waves:} In \gls{BNS} mergers, a part of the mass ejection is driven by a spiral wave arising as a result of the disk-remnant interaction~\citep{spiral2}. Spiral waves in \gls{BNS} mergers---identified by a significant $l=2$, $m=1$ contribution to the gravitational-wave emission---have been studied by~\cite{spiral1, spiral2}. In these studies, it was reported that spiral waves contribute to the ejecta, with the mass calculated using different criteria. The spiral wind ejecta is determined using the Bernoulli criterion once the dynamical ejecta---calculated based on the geodesic criterion---reaches its final value and remains stable. However, in these simulations, the spiral wind and associated ejected matter appear on a longer timescale, around \qty{20}{\milli\second} after the merger.

\cite{spiral1} noted that the spiral  waves continues only as long as the remnant NS does not collapse to the BH, but we note the presence of spiral waves in our \texttt{Q2.6TMAa0} model. In this case, the total mass of the dynamical ejecta reaches its final value around $6.5$ ms after the merger. The mass of the spiral wind ejecta is \qty{1.34e-4}{\Mass\Sun},  reached in \qty{4.3}{\milli\second}. We do not identify the presence of any spiral wind ejecta for the \texttt{Q2.6SFHoa0} model. This suggests a notable dependence on the \gls{EoS}, with stiffer \glspl{EoS} more likely to yield spiral wave-driven ejecta. However, further long-term simulations are needed to validate the actual behaviour of this feature.

\begin{figure}
\includegraphics[width=1.0\columnwidth]{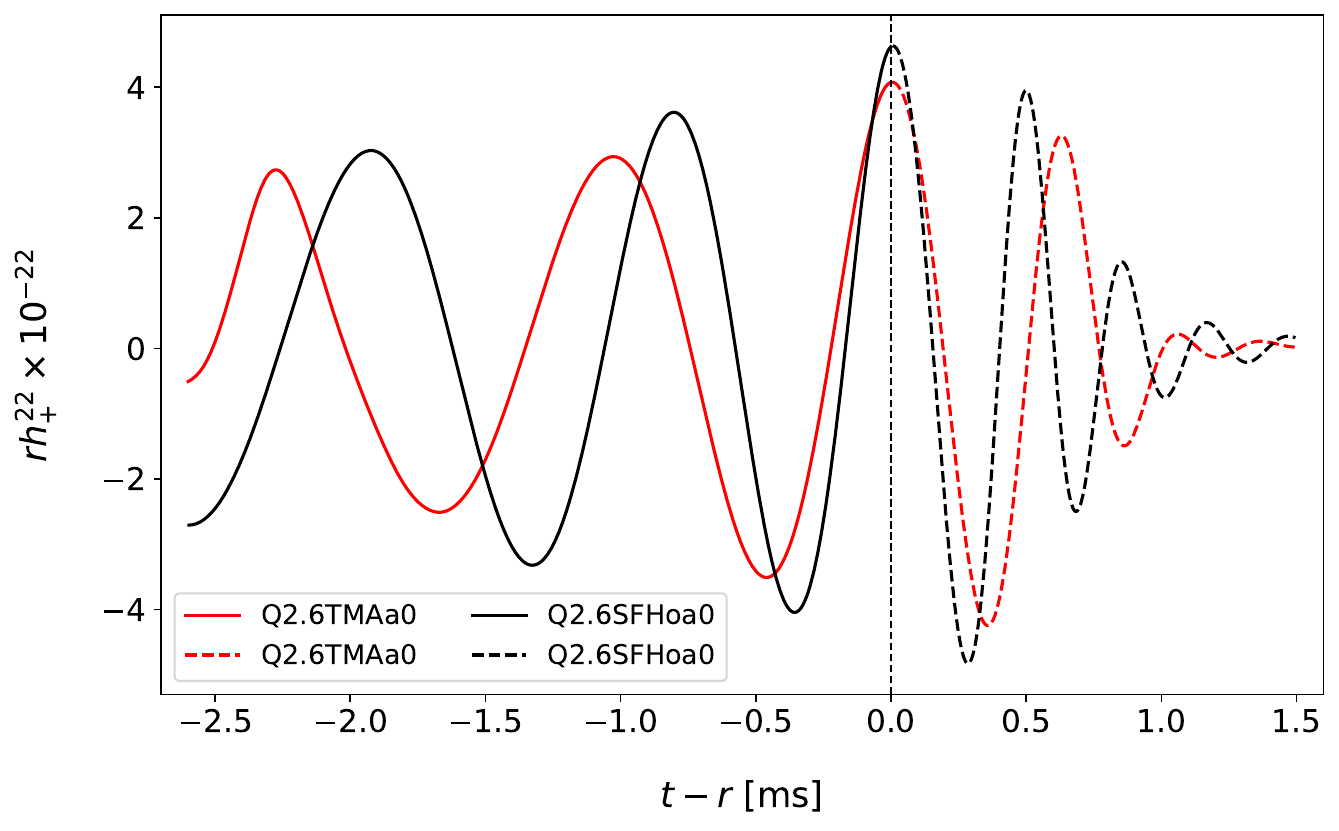}
    \caption{\gls{GW} strain for the $q = 2.6$ mass ratio models. The strains are aligned at the maximum amplitude, which corresponds to the retarded merger time, $t-r=0$, while the retarded time represents the time required for the strain to reach to extraction radius. We extract the strains from a surface located at $\sim \qty{295}{\kilo\metre}$. As can be seen from the figure, the \gls{GW} amplitude is slightly higher for the softer \gls{EoS} in both the inspiral and merger phases.}
    \label{fig:gw230529}
\end{figure}

\textit{Gravitational Waves:} For the gravitational-wave signals we focus on the ability to distinguish the simulated models using observations. To do this, we calculate the signal-to-noise ratio of the dominant $(l, m) = (2, 2)$ mode of the \glspl{GW} for the Advanced LIGO \citep{ligo} and the Einstein Telescope \citep{etelescope} assuming a source distance of $100$ Mpc. The gravitational-wave strains are shown in Figure~\ref{fig:gw230529}. In order to quantify distinguishability, we then work out the minimum required signal-to-noise ratio by first calculating the match between two signals and then identifying the mismatch ($\mathcal{M}$) from the real part of the matched signal. The mismatch is used to find the minimum required signal-to-noise ratio from $\rho_{req} = 1/\sqrt{2\mathcal{M}}$ as described in ~\cite{petergw}. This calculation suggests that the minimum required signal-to-noise ratio to distinguish between the two signals is $\sim2.7$ for both the Advanced LIGO and the Einstein Telescope. As the minimum required signal-to-noise ratio is below the level required to claim a detection in the first place, we expect future gravitational-wave observations of low-mass \gls{BHNS} systems to be able to place useful constraints on the neutron star equation of state.

\textit{Final Properties:} We extract the final properties of the \gls{BH}---quasi-local mass and spin---using the \texttt{QuasiLocalMeasures} (\cite{quasilocal}). For the $q = 2.6$ simulations, we find that the model using the softer \gls{EoS} leaves a \gls{BH}  that is \qty{0.016}{\Mass\Sun} heavier and which has a higher spin. This result is consistent with the disruptive high-mass ratio models discussed in the next section.

\subsection{Impact of the EoS for \texorpdfstring{$q = 5$}{q=5}}\label{section:highmass}

\textit{Dynamics:} We have already discussed the fact that the \gls{EoS} significantly impacts on the merger dynamics for low-mass ratio systems. We now explore this effect for higher mass ratios using a larger set of \glspl{EoS}. As is well known, compact binaries containing aligned spinning \glspl{BH} or \glspl{NS} experience a longer inspiral phase due to the repulsive behaviour of spin-orbit interaction~\citep{campanellispin, 2009PhRvD..79d4024E}. We observe the same behaviour in our simulations: while non-spinning models ($a_{BH} = a_{NS} = 0$) merge approximately in $5$ ms, spinning models (here with $a_{BH} = 0.7$, $a_{NS} = 0$) merge at around \qty{7.5}{\milli\second}.  

Comparing the individual spinning cases we note that, while the \texttt{TMAa07} model merges at \qty{4.2}{\milli\second}, the  \texttt{SFHxa07} case takes \qty{1.2}{\milli\second} longer, merging at \qty{5.4}{\milli\second}. Therefore, even with the stiffest and softest \glspl{EoS} (with a difference in compactness of $\Delta\mathcal{C} \simeq 0.3$), the merger time changes by at most \qty{1.2}{\milli\second} for $q = 5$ and $a = 0.7$.

\glspl{NS} are not expected to be tidally disrupted in high mass ratio, non-spinning \gls{BHNS} mergers. Indeed, as expected from previous simulations, we do not see any tidal disruption for the non-spinning cases, even for the stiffest \gls{EoS} (\texttt{TM10}) model. However, all \glspl{NS} are tidally disrupted for the spinning models at this mass ratio. The question then arises: can we distinguish between different \glspl{EoS} even if the \gls{NS} do not experience tidal disruption? We will present a detailed analysis for this specific mass ratio system to address this question.

 \textit{Mass Ejection:} As can be seen from \cref{table:finalproperties}, with a couple of exceptions (\texttt{TM1a0} and \texttt{TM1a07}), while the total mass ejected matter is not substantial in the non-spinning models (of the order of \qty{e-4}{\Mass\Sun}), both the total mass of the ejected matter and the fast-moving ejecta decrease as the compactness of the \gls{NS} increases. This behaviour is similar to the low mass ratio results. Despite not seeing a clear relationship between the composition of the ejecta and the \gls{EoS}, the electron fraction of the matter remains between $0.037$ and $0.066$,  indicating very neutron-rich matter in all cases. Consequently, \gls{BHNS} mergers can result in heavier r-process nucleosynthesis compared to \gls{BNS} mergers (see, for example, \cite{radicemassejection}). This would, in turn, lead to a dimmer kilonova.

\begin{table*}
\caption{Simulation results at $2.8$~ms after merger. The columns show the final mass of the remnant, the final spin of the \gls{BH}, total ejected mass, electron fraction of the total ejected mass, amount of fast-moving ejecta, the composition of the fast-moving ejecta, the mass of the fallback material and disc.}\label{table:finalproperties}
\small
\begin{tabular}{@{}lSSSSSSSSS@{}}
\hline
		Model & $\unit{\Mass}_{f}[\unit{\Mass\Sun}]$&$a_{f}$&$\unit{\Mass}_{ej}  [\qty{e-4}{\Mass\Sun}]$& $Y_{e,ej}$ & $\unit{\Mass}_{fast}  [\qty{e-5}{\Mass\Sun}]$ & $Y_{e, fast}$ & $\unit{\Mass}_{fb}[\qty{e-5}{\Mass\Sun}]$ &$\unit{\Mass\disk} [\unit{\Mass\Sun}]$\\
		\hline
        Q2.6TMAa0 &4.8540&0.61&6.4&0.032&0.15&0.049&1351.2&0.0888\\
        Q2.6SFHoa0 &4.8880&0.62&2.7&0.042&0.00&0.108&89.3&0.0031\\
  		\hline
		TM1a0 & 8.1198 & 0.440 &    2.5&0.044&2.70& 0.053&8.8& 0.0002\\
        TMAa0 & 8.0090 & 0.446&7.3 & 0.066 & 7.90 & 0.049&15.9&   0.0002 \\
        DD2a0 & 8.0679 & 0.442 & 5.5&0.045&3.40&0.052&7.9&  0.0001\\
        SFHxa0 & 8.0476 &  0.441 & 0.9&0.034&0.70&0.033& 3.7&  0.0000\\
        \hline
        TM1a07 & 7.8852 & 0.819 & 623.4&0.046&1.60&0.050&10272.8& 0.3127\\
        TMAa07 & 7.8687 & 0.815 & 481.2&0.042&3.60&0.053&9893.3& 0.2928 \\
        DD2a07 & 7.9071 & 0.824 & 438.7&0.054&1.00&0.045& 9474.1& 0.1541\\
        SFHxa07 & 7.9854 & 0.833& 254.7&0.043&0.70&0.055&6506.4& 0.0641\\
	    \hline
\end{tabular}
\end{table*}

\begin{table*}
\caption{Final properties of the simulated models \qty{9}{\milli\second} after the merger. Angular distribution calculations show that most parts of the ejecta are within $\pm \ang{30}$ from the equator.}\label{table:final2}
\small
\begin{tabular}{@{}lSSSSSSSS@{}}
\hline
		Model & $\unit{\Mass}_{f}[\unit{\Mass\Sun}]$&$a_{f}$&$\unit{\Mass}_{ej}  [\qty{e-4}{\Mass\Sun}]$& $Y_{e,ej}$ & $\unit{\Mass}_{fast}  [\qty{e-5}{\Mass\Sun}]$ & $Y_{e, fast}$ & $\unit{\Mass}_{fb}[\qty{e-5}{\Mass\Sun}]$ &$\unit{\Mass\disk} [\unit{\Mass\Sun}]$\\
		\hline
        Q2.6TMAa0 &4.8376&0.643&6.8&0.032&0.26&0.043&887.0&0.0638\\
        Q2.6SFHoa0 &4.8576&0.642&2.7&0.042&0.02&0.049&66.7&0.0021\\
        TMAa07 &7.8732&0.844&491.6&0.043&40.00&0.068&8917.0& 0.1570\\
  		\hline
\end{tabular}
\end{table*}

\begin{figure}
\includegraphics[width=1.0\columnwidth]{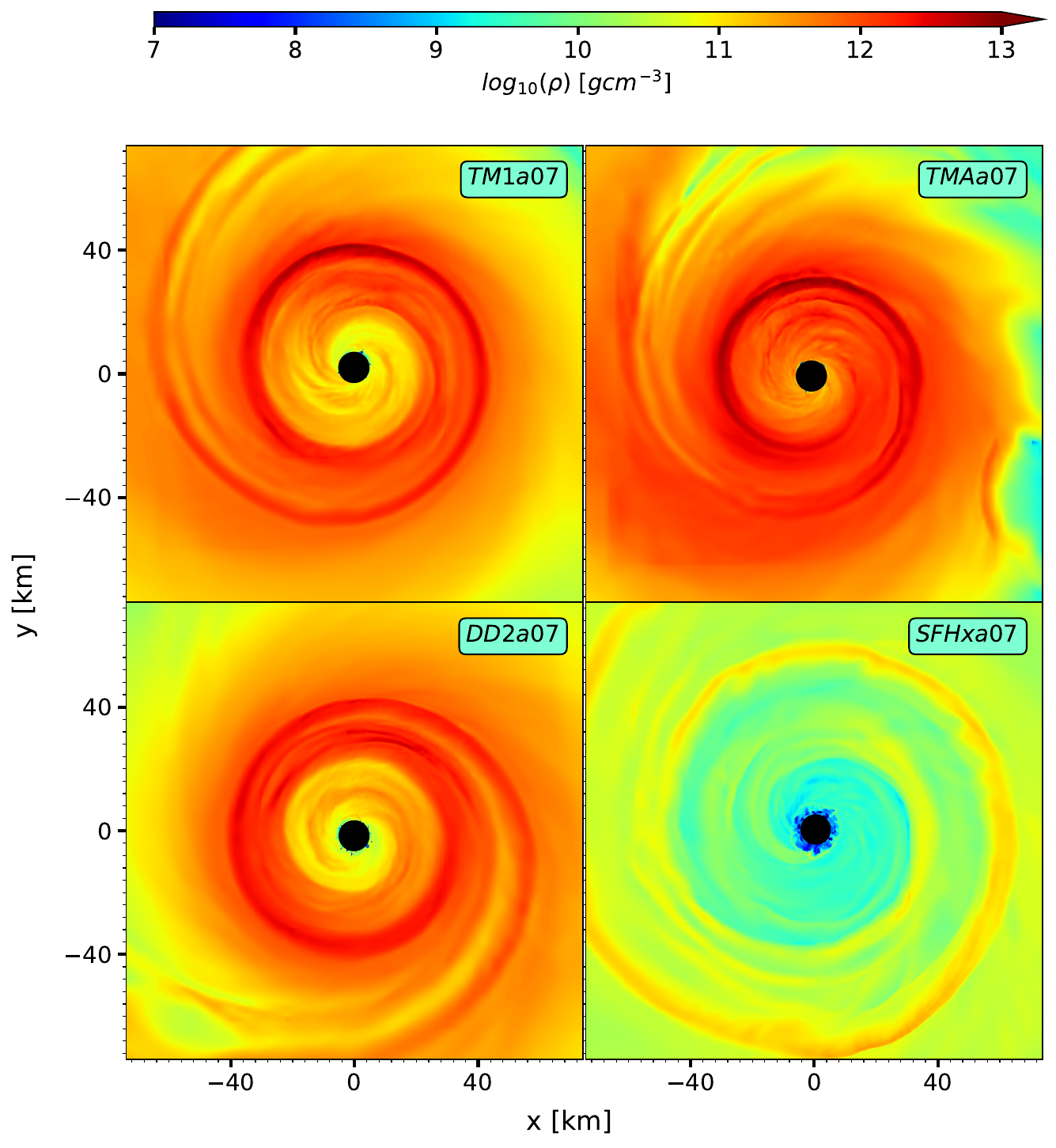}
    \caption{Rest-mass density plot in the equatorial plane for the $q = 5$, spinning BH cases. As can be seen from the results, the matter outside of the \gls{BH} tends to be denser for more compact \glspl{NS}.}
    \label{fig:rest}
\end{figure}

The total ejected mass from the spinning models is at least $65$ times and at most $250$ times greater than that of the corresponding non-spinning models. This implies that, along with different \gls{EoS}, the spin of the \gls{BH}  significantly affects the ejected matter, as expected from previous work (see \cite{kyutokureview} for detailed discussion). The rest-mass density in the equatorial plane for the spinning models is shown in Figure~\ref{fig:rest}.

We also investigated the spiral wave emission for the \texttt{TMAa07} case and found that it never reached the final value of the dynamical ejecta required to determine the corresponding feature. To explore this behaviour in the \gls{BHNS} merger, longer simulations are needed.

 \textit{Comparison with \gls{BNS}:} We also compare our results with irrotational ~\citep{radicemassejection} and highly spinning ~\citep{beyhan} \gls{BNS} mergers. In the case of irrotational configurations, for systems with $\unit{\Mass}_{ADM} = \qty{1.4}{\Mass\Sun}$, ~\cite{radicemassejection} reports that the total ejecta mass for their \texttt{DD2\_M140120\_M0} model is nearly $3$ times higher than in the case of our \texttt{DD2a0} model. This difference is important since unequal mass systems tend to yield higher mass ejection for \gls{BNS} mergers. Even (slightly) unequal mass \gls{BNS} mergers give only 3 times more ejected matter compared to our non-spinning \gls{BHNS} models. Furthermore, the spinning \texttt{DD2a07} model yields approximately $27$ times higher total ejecta compared to the \texttt{DD2\_M140120\_M0} model of \cite{radicemassejection}.

If we compare the spinning \gls{BNS} mergers from~\cite{beyhan} with our spinning \gls{BHNS} mergers, the ejected mass is approximately $1.3$ times higher for a highly spinning equal mass \gls{BNS} ($a_{1,2} = 0.67$) model. Except for this specific model, the spinning \gls{BHNS} models yield more ejected mass than \gls{BNS} mergers. It is, however, notable that~\cite{beyhan} employs a very soft \texttt{SFHo} \gls{EoS}, which is expected to result in more ejected mass for the \gls{BNS} mergers, unlike what is seen for \gls{BHNS} mergers. In general, this suggests that disruptive cases are expected to result in more massive ejecta in \gls{BHNS} mergers.

Furthermore, for high (equal-) mass, irrotational \gls{BNS} models, the composition of the ejected matter has been reported to be very neutron-rich with $Y_{e} = 0.07-0.08$~\citep{radicemassejection}. All other configurations considered in~\cite{radicemassejection} yield neutron-poor ejecta compared to our models. The dynamical ejecta from the highly spinning \gls{BNS} models (individual spins $|a_{1,2}|>0.4$) have nearly the same composition as our models. This behaviour shows that the composition of the ejecta tends to be more neutron-rich in \gls{BHNS} when compared to \gls{BNS} mergers.
 
With the exception of anti-aligned spinning \gls{BNS} mergers with $a=-0.65$, we observe that, in general, \gls{BHNS} mergers for this mass ratio yield slightly higher levels of fast-moving ejecta compared to \gls{BNS} mergers.

\begin{figure*} 
\includegraphics[width=1.0\textwidth]{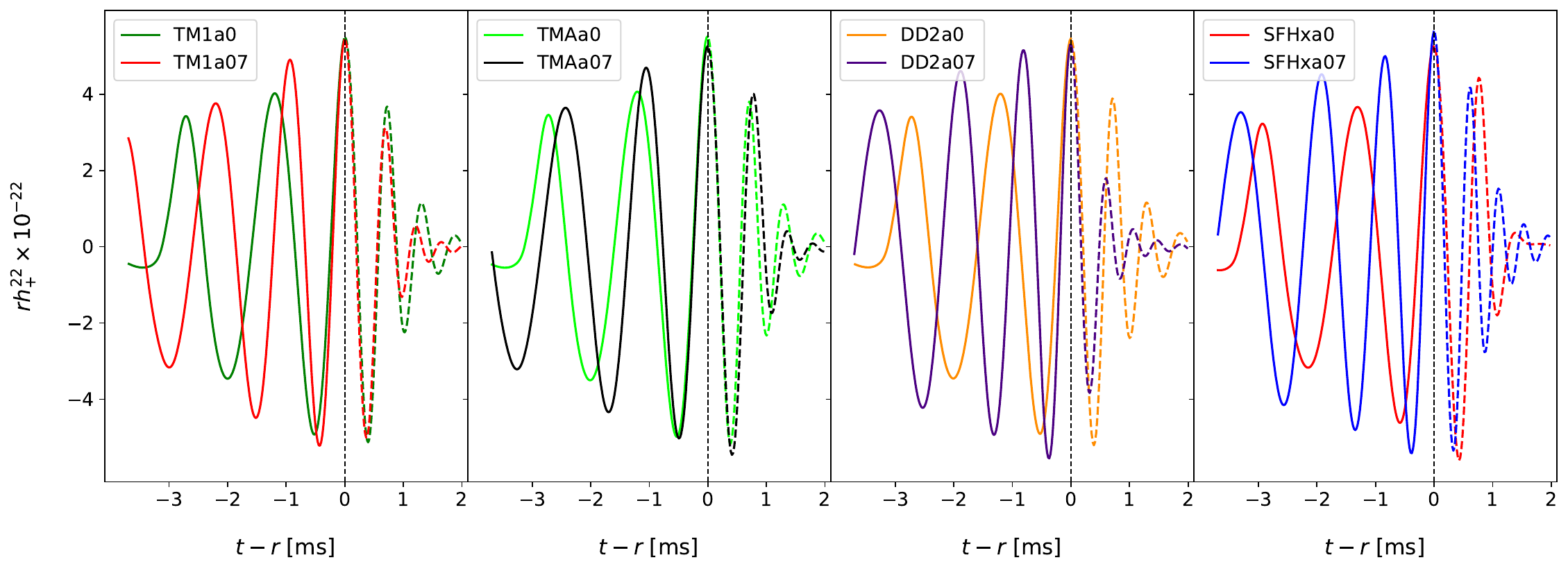}
\caption{\gls{GW} strain for $q = 5$ mass ratio models. Non-spinning and spinning cases are plotted on top of each other, with the \gls{GW} strains aligned at maximum. All \gls{GW} strains are extracted at a distance of $295$ km and the related retarded time has been taken into account. Even though non-spinning models do not show much difference in amplitude, the comparison of the individual spinning models shows that the \gls{GW} amplitude tends to be higher for the softer \glspl{EoS} in both the inspiral and post-merger phases.
}
\label{figure*:gwq8}
\end{figure*}

\textit{Gravitational waves:} The \gls{GW} strain for the $(l,m)=(2,2)$ mode for all cases we have simulated is presented in \cref{figure*:gwq8}, again assuming a source distance of $100$ Mpc. For all models, we extract the \gls{GW} strains at a distance of $\sim \qty{295}{\kilo\metre}$. In \cref{figure*:gwq8}, we plot the non-spinning and spinning models on top of each other, aligning them at maximum to highlight the effect of the spin. We truncate the strain for half an orbit to show the phase difference clearly. Since the non-spinning models merge nearly \qty{2.5}{\milli\second} earlier than the spinning ones, the simulated inspiral is about one orbit shorter for these models.

Again, we determine the distinguishability of the signals using indicative sensitivities for current and next-generation ground-based \gls{GW} detectors, noting that all signals fall within the frequency range where these instruments reach their peak sensitivity. However, given that the sensitivity of current ground-based detectors tends to decrease above a few 100~Hz (due to photon shot noise) and the merger frequency of our models is dominated by higher frequencies, we may require next-generation detectors to observe these merger signals.

We compare the signal-to-noise ratio of the most extreme cases, the softest and the stiffest \gls{EoS} models, for the Advanced LIGO and the Einstein Telescope assuming at a distance of $100$ Mpc. Considering the non-spinning cases, the minimum signal-to-noise ratio required to distinguish these signals are $10.9$ and $12$ for the Advanced LIGO and the Einstein Telescope, respectively.
This suggests,  not surprisingly, that we will need higher signal-to-noise ratio events if we want to constrain the \gls{EoS} for higher mass-ratio systems.

\begin{figure}
\includegraphics[width=1.0\columnwidth]{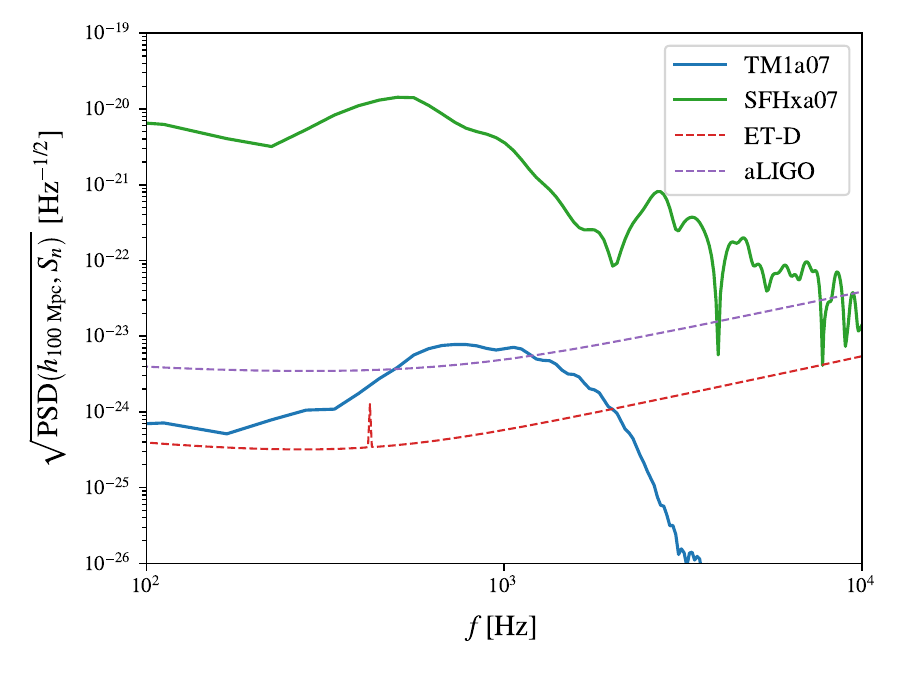}
    \caption{The power spectral density (PSD) of the gravitational-wave signals for the stiffest and softest of the spinning $q = 5$ models, showing the observable difference between the chosen equations of state. While the solid lines show the PSD of the two equations of state models, dashed lines represent the indicative noise curves for the Advanced LIGO and the Einstein Telescope, respectively.}
    \label{fig:psd}
\end{figure}

When we carry out the same analysis for the spinning models, illustrated in~\cref{fig:psd}, we find the minimum signal-to-noise ratios required to distinguish the softest and the stiffest \gls{EoS} models is $0.98$ and $1.01$ using the Advanced LIGO and the Einstein Telescope, respectively. Again, this suggests that the \gls{EoS} can be constrained more confidently for spinning models. 

\begin{figure}
\includegraphics[width=1.0\columnwidth]{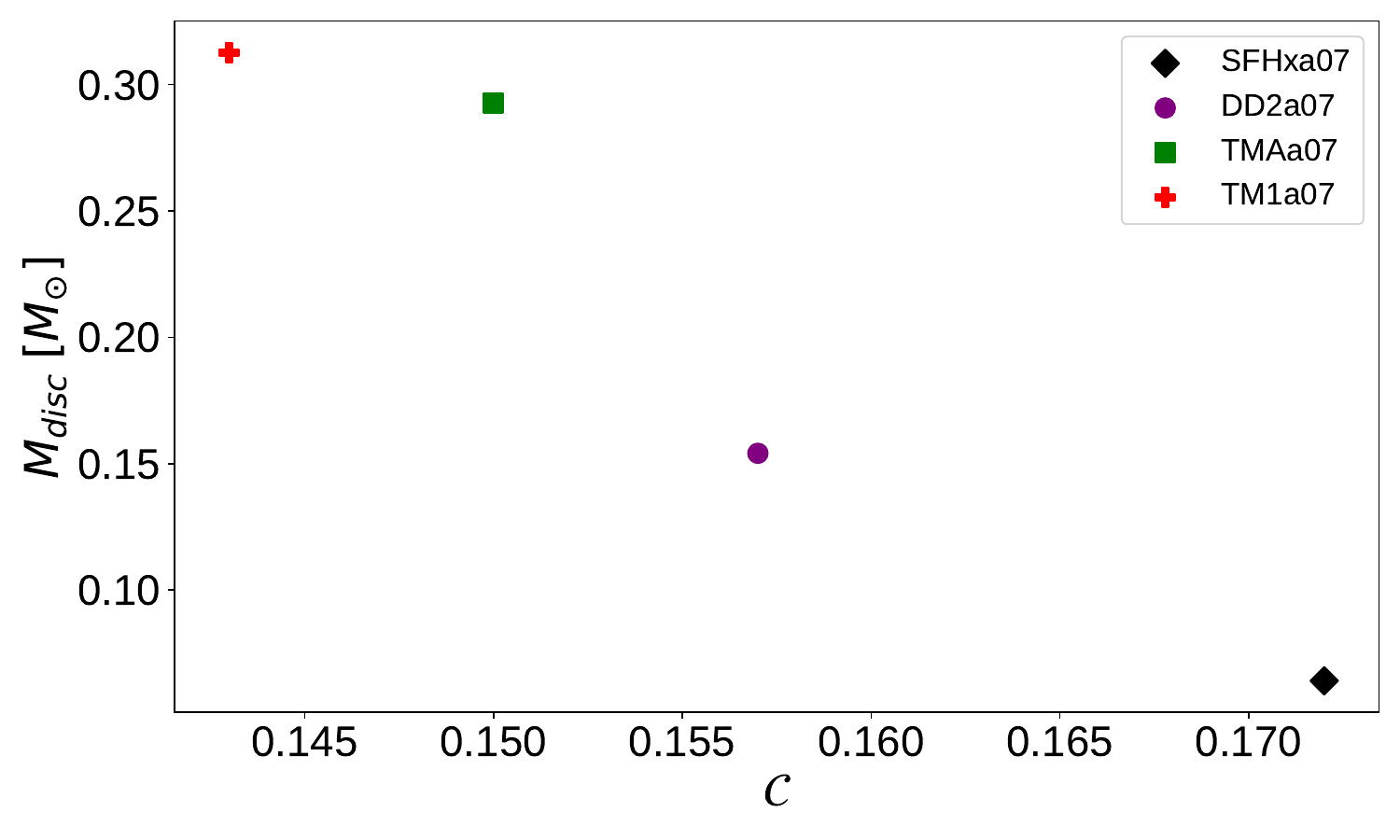}
    \caption{Variation of disc mass with compactness for the $q = 5$ spinning models. Even though our analysis is limited to four \gls{EoS}, that the disc mass increases as the neutron star compactness decreases.}
    \label{fig:disc}
\end{figure}

\textit{Final Mass and Spin:}
Finally, we compare the final mass and spin of the remnant \glspl{BH}. The final properties of all models are provided in \cref{table:finalproperties}. With the exception of the \texttt{TMAa0} case, stiffer \glspl{EoS} tend to produce more massive remnants for non-spinning models. Additionally, except for the \texttt{TM1a0} and \texttt{TMAa07} models, an increase in compactness generally results in a decrease (increase) in the final spin of the black hole for nonspinning (spinning) models.

We present the correlation between the disc mass and compactness for the $q=5$ spinning models in \cref{fig:disc}. Although we only use four \gls{EoS}, the results clearly show that the disc mass increases as the compactness decreases. As the disc mass influences the electromagnetic observables, we suggest that it should be possible to constrain the \gls{EoS}  further through electromagnetic counterparts. Additionally, the correlation seems to hold also for non-disruptive (non-spinning) models, despite these having small disc masses (see \cref{table:finalproperties}). Once further \gls{EoS} are considered, we expect the precise correlation to become clearer.

\subsection{Impact of the Mass Ratio}

In order to explore the impact of the mass ratio, we focus on the ejecta properties for three models: \texttt{Q2.6TMAa0}, \texttt{TMAa0}, and \texttt{TMAa07}. We end the evolution of the \texttt{TMAa0} model at $t - t_{\text{merger}} = \qty{5}{\milli\second}$, since the maximum rest-mass density in the domain then reduces to \num{e4}$ \rho_{atmo}$. The other two models were evolved until \qty{10}{\milli\second} after merger. We compare \texttt{Q2.6TMAa0} and \texttt{TMAa0} at \qty{2.8}{\milli\second} ms after merger, while comparing \texttt{Q2.6TMAa0} and \texttt{TMAa07} at \qty{9}{\milli\second} after merger.

The comparison of \texttt{Q2.6TMAa0} and \texttt{TMAa0} illustrates the dependence of the mass ratio on the mass of the total ejecta and its fast-moving component. For the high mass ratio model, the mass of these two components increases by a factor of  $1.14$ and $50$, respectively. However, the mass of the fallback material does not show the same dependence. It is $85$ times less massive in the high mass ratio model compared to low mass ratio models. Despite the amount of total ejecta being nearly the same, its composition becomes half as neutron-rich in the high mass ratio model. This difference in the composition will be reflected in the electromagnetic counterpart of these events. Furthermore, whilst the mass of the fast-moving ejecta significantly increases for the high mass ratio model, the composition does not change considerably with the mass ratio.

Even though the \texttt{TMAa07} model has the spin of $a = 0.7$ and the \texttt{Q2.6TMAa0} model is non-spinning, we compare them as disruptive \gls{BHNS} mergers at \qty{9}{\milli\second} after merger. At this time, in the spinning model---while the mass of the fallback material is $\sim 10$ times higher---the total ejecta dramatically increases and becomes $72$ times more massive as compared to the non-spinning model. Furthermore, the disk of the \texttt{TMAa07} model is nearly $2.5$ times more massive than in the case of the \texttt{Q2.6TMAa0} model.

Investigation of selected mass ratio cases shows that spin has a greater impact on tidal disruption than the mass ratio, and spinning binaries experience more violent disruptions considering the dynamical ejecta and fallback material.

\section{Summary and Conclusions}\label{sec:conclusion}

In this study, we have investigated---using numerical relativistic simulations---the conditions under which equation of state information may be extracted from black hole-neutron star mergers. We performed ten simulations for two mass ratios, $q = 2.6$ and $q = 5$, five equations of state (ranging from soft to stiff) and two spin configurations for the BH, $a = 0$ and $a = 0.7$. This extends previous work by investigating a broader range of equations of state. For the $q = 2.6$ models, we targeted the recently observed  GW230529 event and discussed whether we should expect to see any differences between two equation of state models. In addition, we investigated the effect of the mass ratio and the spin for $q = 5$ models. 

Our results show that the total mass of ejected matter and the fast-moving ejecta component both increase as the compactness of the NS decreases, regardless of the black hole spin and the mass ratio. However, in the low mass ratio systems the effect of the matter equation of state is much more prominent. The stiff equation of state model experiences more violent tidal disruption by producing a much more massive fallback material and remnant disc as compared to the soft equation of state. The fact that the stiff equation of state produces a disc with a mass of $\qty{0.07}{\Mass\Sun}$ unlike the other model,  will most likely be distinguishable through electromagnetic counterparts.

Furthermore, even though our high mass ratio, non-spinning models do not show tidal disruption, the amount of ejected matter is not negligible. In fact, it becomes more massive compared to the low mass ratio, disruptive model. In addition to this, the fast-moving ejecta is more massive for non-disruptive high mass ratio models, except for the case of the \texttt{SFHx} equation of state. This suggests that the fast-moving component of the ejecta could be used to distinguish the equation of state even for non-disruptive models (at least for this mass ratio, $q = 5 )$.

Even though we studied only five equations of state models, our results indicate that there is a correlation between disc mass and the compactness of neutron stars in black hole-neutron star mergers. The relationship holds for both low- and high-mass ratio models, as well as spinning and non-spinning models. A larger set of models is required to work out the precise correlation, we suggest that the equation of state may also be constrained using kilonova observations.

The comparison of disruptive models with low and high mass ratios reveals that the mass of the ejecta, fallback material, and disc are much higher for the latter. This shows that the spin has greater impact on tidal disruption than the mass ratio for these configurations.

The composition of the ejected matter is highly neutron-rich for all cases we have studied, suggesting heavier r-process nucleosynthesis and consequently dimmer kilonovae for black hole-neutron star mergers compared to binary neutron star mergers. However, we do not see a clear relationship between the composition of the ejecta and the equation of state.

We noticed the presence of a spiral wave in the \texttt{Q2.6TMAa0} model and calculated the associated mass of the ejecta for the \texttt{Q2.6TMAa0}, \texttt{Q2.6SFHoa0}, and \texttt{TMAa07} at \qtylist{9}{\milli\second} after merger. Even though we only see the presence of a spiral wind ejecta for the \texttt{Q2.6TMAa0} model---a low mass ratio, stiff equation of state model---longer-term simulations are needed to better understand the importance of this ejecta component and its electromagnetic counterpart for black hole-neutron star mergers. This is an interesting issue, but we leave it for future work.

We investigated the gravitational-wave distinguishability of the equation of state by calculating the signal-to-noise ratio of all models using the Advanced LIGO and the Einstein Telescope sensitivities, assuming a distance of $100$ Mpc. Based on these mismatch calculations, we do not expect to easily constrain the equation of state using gravitational waves for high mass ratio, non-spinning models using Advanced LIGO. However, the \texttt{TM1a0} and \texttt{SFHxa0} models---corresponding to the stiffest and the softest equation of state models we considered---should be easily distinguished for the high signal-to-noise events expected with next-generation ground-based gravitational wave detectors like the Einstein Telescope.  

To sum up, we conclude that it should be possible to distinguish the equation of state of the neutron star and constrain the initial black hole spin using observed gravitational wave signals and the ejected matter properties for disruptive black hole-neutron star mergers. In future work, it would be interesting to investigate further the possible electromagnetic counterparts using r-process nucleosynthesis for finite-temperature, composition-dependent equations of state models with the inclusion of realistic magnetic field and neutrino treatments. 

\section*{Acknowledgements}

RM would like to thank Beyhan Karakas and Bruno Giacomazzo for their support, Wolfgang Kastaun for sharing the extra analysis tools and \texttt{Spritz} group for useful discussions. We acknowledge the use of the IRIDIS High Performance Computing Facility, and associated support services at the University of Southampton. NA and IH also gratefully acknowledge support from the Science and Technology Facility Council (STFC) via grant numbers ST/R00045X/1 and ST/V000551/1.

\section*{Data Availability}

 An example of the parameter file and ID will be published on Zenodo and \gls{GW} strains will be uploaded to the \texttt{CoRe} database \cite{core}.

\bibliographystyle{mnras}
\bibliography{bhns}

\bsp
\label{lastpage}
\end{document}